\documentclass[reprint, aps, prd, floatfix,nofootinbib,letter, superscriptaddress, nodate]{revtex4-2}
 
\usepackage[T1]{fontenc}
\usepackage[utf8]{inputenc}
\usepackage{lmodern}
\usepackage{setspace}
\usepackage[marginal, multiple]{footmisc}
\usepackage[colorlinks, allcolors=blue!70!black, linktocpage]{hyperref}

\usepackage{amsmath, amssymb, amsfonts, amsthm, latexsym, mathrsfs, xcolor, bbm, slashed, braket, cancel, graphicx, mathtools, setspace, orcidlink, enumitem}

\newcommand{\scrip}{\mathscr{I}^+}
\newcommand{\scrim}{\mathscr{I}^-}

\newcommand{\be}{\begin{equation}\begin{aligned}}
\newcommand{\ee}{\end{aligned}\end{equation}}
\newcommand{\ra}{\rightarrow}

\usepackage[allcommands,old-arrows]{overarrows}

\begin{document}

\count\footins = 800 

\title{Minimum lifetime of a black hole}

\author{Eugenio Bianchi \orcidlink{0000-0001-7847-9929}}
\email{ebianchi@psu.edu}
\affiliation{Institute for Gravitation \& the Cosmos\\ and Department of Physics, The Pennsylvania State University\\ University Park, PA 16802, USA}

\author{Matthew Brandsema \orcidlink{0000-0002-4530-5946}}
\email{mjb619@arl.psu.edu}
\affiliation{Applied Research Lab, The Pennsylvania State University\\University Park, PA 16802, USA}

\author{Kenneth Czuprynski \orcidlink{0009-0005-6342-8420}}
\email{kdc168@psu.edu}
\affiliation{Applied Research Lab, The Pennsylvania State University\\University Park, PA 16802, USA}

\author{Daniel E. Paraizo \orcidlink{0000-0002-8653-0971}}
\email{dparaizo@psu.edu}
\affiliation{Institute for Gravitation \& the Cosmos\\ and Department of Physics, The Pennsylvania State University\\ University Park, PA 16802, USA}
\affiliation{Applied Research Lab, The Pennsylvania State University\\University Park, PA 16802, USA}


\begin{abstract}

We derive bounds on the lifetime of an evaporating black hole. The minimum lifetime bound follows from energy conservation and purification, within the framework of `asymptotically semiclassical spacetimes'. The recently derived expression for the Bondi flux of Hawking radiation, together with the expression for the entanglement entropy of Hawking radiation at null infinity, is used to investigate the purification phase after the last semiclassical ray. We discuss the energy-cost of entanglement purification and find a lower bound on the purification time of the black hole, which scales as $M_0^4/\hbar^{3/2}$, where $M_0$ is the initial black hole mass. Additionally, motivated by the Planck luminosity bound in quantum gravity, we include the additional assumption that a Planck mass black hole is metastable. With this assumption, the purification time is in fact exponential in the square of the initial black hole mass, $\tau\sim\sqrt{\hbar}\,\exp(\gamma M_0^2/\hbar)$ with $\gamma>0$, i.e., exponential in its initial area. We find that the redshift exponent is negative in this purification phase, which indicates the existence of a white-hole remnant that releases information slowly. We comment on phenomenological implications for primordial black hole remnants.
    
\end{abstract}

\maketitle


\section{Introduction}

In the initial adiabatic phase of evaporation that should be well-described by semiclassical physics, it takes a black hole a time $M_0^3/\hbar$ to evaporate from its initial mass $M_0$, down to a few Planck masses \cite{Hawking:1974rv,Hawking:1975vcx, Page:1976df}. During this phase, the Hawking radiation in the exterior of the black hole will be in a mixed state \cite{wald_1975, Hawking1976} (see also \cite{BirrellandDavies, Wald_QFTCS, FabbriandSalas, Hossenfelder_2010, Unruh_2017, Marolf_2017, Ashtekar_2020} for excellent reviews and perspectives). If information is recovered, how much longer will it take for the black hole to purify?

In this paper, we show that energy conservation and unitary evolution provide strong constraints on the purification time of a black hole. In particular, by combining the expressions for the radiative flux of Hawking radiation found in \cite{BianchiParaizo2026} with the entanglement entropy of radiation found in \cite{Bianchi:2014qua,Bianchi:2014vea,Bianchi:2014bma}, we can provide a remarkably simple proof that the minimum lifetime of a black hole is $M_0^4/\hbar^{3/2}$. Such bounds have been qualitatively explored in, e.g., \cite{Carlitz-Wiley1987_Lifetime, preskill1992blackholesdestroyinformation, phdthesis}. Moreover, under an additional physical assumption of metastability, together with the constraints of energy-conservation and purification, we find a configuration with an exponential lifetime of $\sqrt{\hbar} \,\mathrm{e}^{\gamma M_0^2/\hbar}$, where $\gamma>0$, which results in a metastable Planckian remnant with negative redshift exponent, similar to the ones investigated in \cite{Bianchi:2018mml}.

To investigate these issues, we consider the scenario of  `asymptotically semiclassical spacetimes'. The essence of this framework is that one assumes quantum-gravitational effects dissipate far from an evaporating black hole, and imposing the semiclassical Einstein equation asymptotically amounts to taking expectation values of operator-valued balance laws at null infinity. We investigate the consequences of this scenario in the standard spherically-symmetric model of black hole evaporation in $4d$, and illustrate it with its moving mirror analogue \cite{Davies-Fulling_MM, Davies-Fulling_MM+BH, Davies-Fulling76,Varadarajan-2025}.

Within this model, we further obtain an energy-entropy relation and investigate its consequences. In the scenario considered, the curve for the entanglement entropy of radiation \cite{Page_1993_EE, Page_1993_RadInfo, Page_2013} has a turning point at the Planck mass, which corresponds to the `last semiclassical ray'. After this point, the correlations in the Hawking radiation at null infinity cannot be purified at zero energy cost \cite{Hotta_PartnerParticles, Wald_2019, Agullo:2026}. The results provide new bounds on the lifetime of the resulting remnants \cite{AHARONOV198751, Banks:1992is, Giddings:1993vj, Giddings:1993km, Banks:1994ph, Adler:2001vs, Chen:2014jwq, Christodoulou_2016, Bianchi:2018mml}, and can have phenomenological consequences for primordial black holes \cite{Vidotto:2026pbh}.

Throughout the paper we do not report factors of $c$, $G_N$, $k_B$ which can be easily restored, but keep the Planck constant $\hbar$ to tell apart classical from quantum phenomena. With these conventions, the Planck mass is $m_P=\sqrt{\hbar}$ and the Planck area is $a_P=\hbar$.


\section{Asymptotically ${}$ semiclassical\\[.2em] spacetimes}

In classical general relativity, null infinity serves as a natural arena to describe asymptotic observables of isolated gravitational systems. The radiative modes are captured by the fluxes $\mathcal{F}_\xi$ associated with BMS vector fields $\xi$ \cite{Geroch1977, Ashtekar-Streubel, Ashtekar_AQ, Ashtekar:1987tt, Ashtekar:2014zsa, Ashtekar:2018lor, Bonga_2020, Ashtekar_2024, ParaizoWald}. As emphasized by Ashtekar and Speziale in \cite{Ashtekar_2024}, one only needs the radiative phase space to identify the fluxes, without the need of full knowledge of field equations or the covariant phase space. In particular, one does \textit{not} need to solve the Einstein equations in the spacetime bulk. In this sense, future null infinity $\scrip$ is treated as an `abstract manifold', and one can determine fluxes for \textit{all} asymptotically flat spacetimes, regardless of what processes occur in the bulk spacetime. 

Charges $\mathcal{Q}_\xi$, on the other hand, do require field equations and the full covariant phase space. \textit{Thus, they contain more information than what the radiation at null infinity carries.} Indeed, charges can be regarded as `boundary data', defined on $2$-sphere cross-sections of $\scrip$, and hence are of Coulombic nature. The change in the charge $\mathcal{Q}_\xi$ from the cross-sections $S_1$ to $S_2$ is given by the flux $\mathcal{F}_\xi$ through the compact portion $\Delta \scrip$ of null infinity which they bound. This relation is the asymptotic balance law $\mathcal{Q}_\xi[S_2] - \mathcal{Q}_\xi[S_1] = - \mathcal{F}_\xi[\Delta \scrip]$.  

\smallskip

The quantum theory can also be formulated intrinsically at future null infinity by focusing on asymptotic observables. In Ashtekar's asymptotic quantization program \cite{Ashtekar_AQ, Ashtekar:1987tt, Ashtekar:2014zsa, Ashtekar:2018lor}, one introduces an asymptotic Hilbert space of states, $\mathcal{H}_{\mathrm{out}}$, which is given by the Fock space of perturbative radiative quantum fields. The classical balance laws at $\scrip$ then result in operator equations
\begin{equation} \label{balance_law}
    \hat{\mathcal{Q}}_\xi[S_2] - \hat{\mathcal{Q}}_\xi[S_1] = - \hat{\mathcal{F}}_\xi[\Delta \scrip] \,.
\end{equation}
Given a quantum state $\ket{\Psi}$ at future null infinity, the expectation value of the balance law \eqref{balance_law} provides the asymptotic version of the semiclassical Einstein equations. 

In the formation and evaporation of a black hole, the spacetime geometry in the bulk reaches Planckian curvature and quantum geometry effects cannot be neglected. However, as the system is isolated, asymptotic observables should be able to be treated perturbatively once we are given the quantum state $\ket{\Psi}$ at future null infinity. Thus, the balance law \eqref{balance_law} would be satisfied, even though gravity is not semiclassical in the bulk.
Clearly, predicting the state $\ket{\Psi}$ from initial data requires control of full quantum gravity. For instance, one would need to know how to map the in-vacuum $\ket{0_\mathrm{in}}\in \mathcal{H}_{\mathrm{in}}$ to a state in $\mathcal{H}_{\mathrm{out}}$. Instead, in this paper, we will work purely on $\scrip$: We consider only the out-Hilbert space, and identify constraints on the possible quantum states that arise by imposing the quantum balance laws \eqref{balance_law}, together with purity of the state $\ket{\Psi}\in \mathcal{H}_{\mathrm{out}}$.

We consider first the energy flux relevant for black hole evaporation. It is important to note that here we focus on the energy flux at null infinity, not the flux through the horizon (which we have no access to in the asymptotic framework). Indeed, the horizon flux can include massive particles which a far-away observer would never detect. Instead, such `Coulombic' contributions should be regarded as a `thermal cloud' surrounding the black hole, and hence as part of the mass measured by an asymptotic observer. This perspective was adopted in \cite{BianchiParaizo2026} where the expression for the radiative energy flux for a massless minimally-coupled scalar field was first determined at the classical level in (3+1) dimensions using the generalized Wald-Zoupas prescription for fluxes at null infinity \cite{Wald_2000, Harlow:2019yfa, Bonga_2020, Grant:2021sxk, Odak:2022ndm, Odak:2023pga, Rignon-Bret:2024gcx, Rignon-Bret-Speziale2025}. The fluxes in the quantum theory followed from the quantization of these classical expressions. In particular, the Bondi mass $M(u)$ is then \textit{inferred} from the flux and the balance law \eqref{balance_law}, and is given by
\begin{equation}
\label{mass-balance}
M(u)=M_0-\int_{-\infty}^u \langle F_{\mathrm{rad}}\rangle(u)\,.
\end{equation}
Here, $u$ is the retarded time at $\scrip$, the formal expression $\langle F_{\mathrm{rad}}\rangle(u)$ is the renormalized expectation value of the radiative flux operator on any suitable state $\ket{\Psi}\in \mathcal{H}_{\mathrm{out}}$, and $M_0=M(-\infty)$ is the initial ADM mass of the black hole. 

Adopting the usual spherically-symmetric $s$-wave model for black hole evaporation \cite{Davies-Fulling_MM, Davies-Fulling_MM+BH, Davies-Fulling76,Varadarajan-2025}, the relevant state-space is reduced to conformal vacua and thus the correlation functions in $\langle F_{\mathrm{rad}}\rangle(u)$ can be easily computed. In this spherical model the radiative gravitational degrees of freedom are identically zero. As such, in the remainder of the paper we need not worry about the infrared sector of the gravitational field. The radiative energy flux for the scalar field is given by \cite{BianchiParaizo2026}
\begin{equation} \label{rad_flux}
    \braket{F_{\rm rad}}(u) =\, \alpha\, \hbar\, k(u)^2,
\end{equation}
where $\alpha$ is a constant that depends on the type of field under consideration. Specifically, for the $s$-wave of $N$ massless scalar fields, the constant is 
\begin{equation}\label{alpha}
\alpha = \frac{N}{48 \pi}\,.
\end{equation}
The function  $k(u)$ is the redshift exponent \cite{Barcel2011-HawkRad, Barcel2011-Minimal_Conditions, Agullo_2024, Agullo:2026} which is encoded in the correlation functions in the state $|\Psi\rangle\in \mathcal{H}_{\mathrm{out}}$ and does not require knowledge of the bulk spacetime.

Next, we consider the entropy of radiation, $S_{\mathrm{rad}}(u)$. Again, this quantity is defined intrinsically at $\scrip$: it depends only on the state $\ket{\Psi}$ and a cross section $S$ labeled by retarded time $u$. It is defined as the renormalized entanglement entropy  due to correlations between the early portion and the \emph{late} portion $(u,+\infty)$ of $\scrip$. Within the same spherically-symmetric model for black hole evaporation \cite{Davies-Fulling_MM, Davies-Fulling_MM+BH, Davies-Fulling76,Varadarajan-2025} used for the flux \eqref{rad_flux}, the entropy of radiation could be computed for conformal vacua and was found to be completely determined by the redshift exponent $k(u)$. It is given by the expression \cite{Bianchi:2014bma, Bianchi:2014qua, Bianchi:2014vea}:
\begin{equation} \label{EE_formula}
    S_{\mathrm{rad}}(u) = 4 \pi \alpha \int_{-\infty}^u k(u')\, du'.
\end{equation}
The renormalized entropy of radiation represents excess entropy compared to that of the vacuum $\ket{0_{\mathrm{out}}}$. 

\medskip

The asymptotic approach adopted in this paper allows us to determine bounds on the lifetime of a black hole, stemming from considerations of energy \eqref{rad_flux} and entanglement \eqref{EE_formula}, without determining the quantum geometry in the bulk. This analysis depends on two principle assumptions regarding a quantum theory of gravity, whose implications we go over in more detail in the Discussion section. However, for clarity, we spell them out now. We \textit{assume} that:
\begin{itemize}
    \item[(i)] Planckian curvature does not propagate out to infinity, i.e., there is a notion of `\textit{asymptotically semiclassical spacetimes}'. This is meant only with respect to expectation values, and \textit{not} fluctuations in the curvature or the stress tensor \cite{FordKuo_Semicl, Wu_1999, Ford:2000vm};

    \item[(ii)] Information is recovered at future null infinity \cite{Page_1993_EE, Page_1993_RadInfo, Page_2013}, so that the entanglement entropy of radiation at infinity will vanish in the far-future. Note we do not assume or impose the condition that the entropy of radiation turns overs at the Page time, i.e., when its value matches the value of the Bekenstein-Hawking horizon entropy. The entanglement entropy of radiation is purely defined at future null infinity, and its turn-over time is determined by the state which is assumed to be pure.
\end{itemize}

\begin{figure}
  \centering
  \includegraphics[scale=0.65]{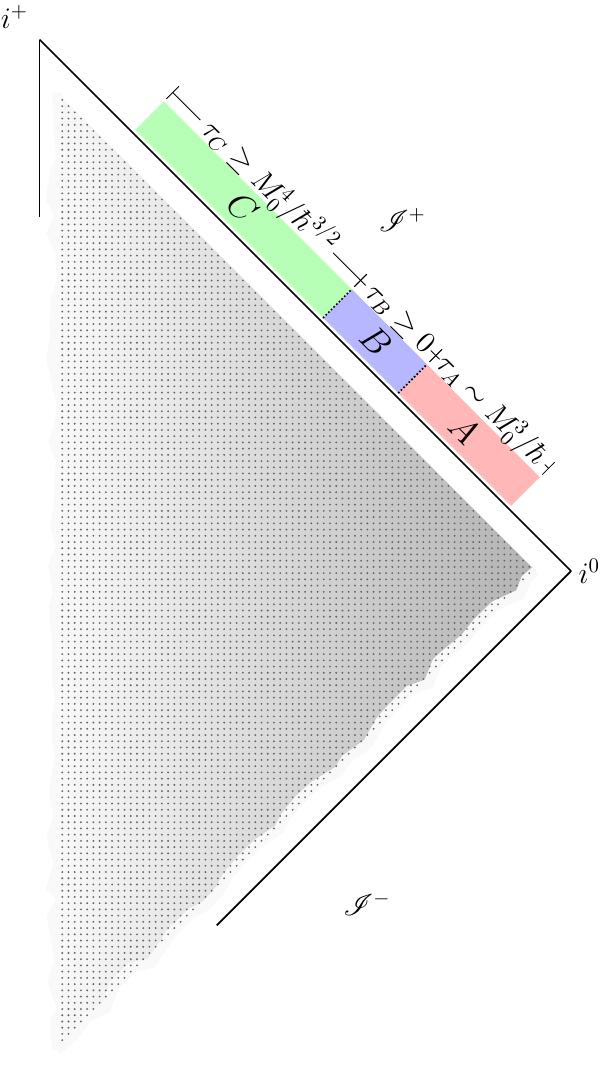}
  \caption{The Penrose diagram for an evaporating black hole. The shaded grey region represents the bulk which is left arbitrary. The colored asymptotic regions represent the three different phases of evaporation. Assuming information is recovered, we constrain how long the purification phase C must last.}
  \label{penrose_diagram}
\end{figure}

\section{Phases of Evaporation}

Under these assumptions, at future null infinity, the process of black hole evaporation can be neatly split into three phases, as described in Fig.~\ref{penrose_diagram}. \emph{Phase A} is the initial Hawking phase, where thermal Hawking radiation is emitted adiabatically by the black hole. This phase is expected to be well-described by semiclassical gravity, both in the bulk and at $\scrip$. As a result, the redshift exponent $k_A(u)$ can be approximated directly by the surface gravity \cite{Barcel2011-HawkRad, Barcel2011-Minimal_Conditions, Agullo_2024, Agullo:2026}, i.e.,
\begin{equation}
\label{kA}
k_A(u)=\frac{1}{4M(u)}\,,
\end{equation} 
where $M(u)$ is the Bondi mass at the retarded time $u$. This phase ranges from the time $u_0$ when the black hole initially forms, to the `last semiclassical ray' $u_{\mathrm{lr}}$ when the black hole reaches the Planck mass $M(u_{\mathrm{lr}})\approx \sqrt{\hbar}$. The estimate of the duration $\tau_A \equiv u_{\mathrm{lr}}-u_0$ of this phase is standard and immediate: one solves the balance law \eqref{mass-balance} which takes the form $\partial_u {M}(u)=-\alpha \hbar\, (4M(u))^{-2}$, and finds
\begin{equation}
\label{tauA}
\tau_A \approx\frac{16}{3\alpha} \frac{M_0^3}{\hbar}\,.
\end{equation}
This is Hawking's celebrated result \cite{Hawking:1974rv, Hawking:1975vcx}. During this phase, the mass decreases and the entropy of radiation increases. The question we address here is what happens next, after we exit the regime of phase A where semiclassical gravity applies in the bulk.

\medskip

\emph{Phase B} is a transition period of possible quiescence, which we define as a phase with vanishing redshift exponent,
\begin{equation}
k_B(u)=0\,.
\end{equation}
In this phase, the radiative flux \eqref{rad_flux} vanishes and therefore the mass remains Planckian. Moreover, the radiation entropy \eqref{EE_formula} remains constant. Conventionally, we set the end of this phase at the retarded time $u=0$. Predicting the existence and duration $\tau_B \equiv | 0-u_{\mathrm{lr}} |$ of this phase requires a full theory of quantum gravity. Here we will simply consider its possible existence, without putting a bound on its duration: 
\begin{equation}
\tau_B\geq 0\,.
\end{equation}

\emph{Phase C} is the purification phase. This phase also requires a full theory of quantum gravity for predicting its detailed properties. However, here, we are interested only in a lower bound on its expected duration $\tau_C$. In the next section we will show how this bound follows from constraints on purity and energy conservation.


\medskip

To clarify the properties of the state $\ket{\Psi}$ in the three phases $A$, $B$, and $C$, it is useful to discuss correlation functions for perturbative radiative quantum fields, specifically the scalar field $\hat{\chi}(u,\theta,\phi)$.  
In particular, we can distinguish between three classes of two-point functions for the scalar field $\chi$ which is responsible for Hawking radiation in the spherically-symmetric model. First, $G_{AA} = \bra{\Psi} \chi(u_A, \theta, \phi) \chi(u'_A, \theta, \phi) \ket{\Psi}$ describes the correlations in the radiation between two points $u_A$ and $u'_A$ that lie in the spacetime region corresponding to phase $A$, and as such is thermal at the Hawking temperature. On the other hand, $G_{AC}$ and $G_{CC}$ respectively describe correlations in the radiation between points $(u_A, u_C)$ in the spacetime regions corresponding to phases $A$ and $C$, and points $(u_C, u'_C)$ in the spacetime region corresponding to phase $C$ only. Remarkably, for the $(3+1)$d spherically-symmetric model of evaporation \cite{Davies-Fulling_MM, Davies-Fulling_MM+BH, Davies-Fulling76,Varadarajan-2025}, the combination of two-point functions that enters into the energy flux \eqref{rad_flux} and entanglement entropy \eqref{EE_formula}, is completely determined by a single parameter, the redshift exponent $k(u)$. This occurs, in essence, due to our restriction to conformal vacua. Therefore, varying over the possible two-point functions (and hence, quantum states within the conformal-vacuum family) amounts to varying over the single function $k(u)$. In the next section, we will use this property to bound the duration of the purification phase $C$.


\section{Constraints on Evaporation}

At the beginning of phase $C$, the Bondi mass is Planckian, $M(0)=\sqrt{\hbar}$, where we have set conventionally the beginning of phase $C$ to the retarded time $u=0$. The first constraint we impose on the purification phase is that, at its end, the Bondi mass vanishes. Using the balance law $\eqref{mass-balance}$ with the radiative flux \eqref{rad_flux} and the conditions
\begin{equation}
M(0)=\sqrt{\hbar}\,,\qquad M(\tau_C)=0\,,
\label{eq:planck-mass}
\end{equation}
this determines the first constraint
\begin{equation} \label{energy_constraint}
    \int_0^{\tau_C} k(u)^2\, du = \frac{1}{\alpha\, \sqrt{\hbar}},
\end{equation}
which can be understood as an energy-budget constraint.

Next, we consider the entropy of radiation. As discussed, phase $A$ corresponds to the emission of thermal Hawking radiation, so that the entanglement entropy increases up to the last semiclassical ray where it reaches the leading-order value $S_{\mathrm{rad}}(u_{\mathrm{lr}}) =  A_0/ 2 \hbar$ where $A_0=16\pi M_0^2$ is the initial horizon area. This expression can be found directly using \eqref{EE_formula} by integrating the redshift exponent \eqref{kA} over the time $\tau_A$ \cite{Bianchi:2014bma}. The factor of $2$ ``discrepancy'' with respect to the Bekenstein-Hawking formula for the entropy of the black hole \cite{Hawking:1974rv, Bekenstein73, Hawking1976_BHT, AshtekarParaizoShu-PRL, AshtekarParaizoShu-GRG} is due to evaporation being an irreversible process \cite{Zurek1982, Bianchi:2014bma}.

Now, combining the expressions for the radiation entropy  \eqref{EE_formula} and the radiative flux \eqref{rad_flux} yields the energy-entropy relation:
\begin{equation} \label{energy_entropy_rel}
    \dot{M}(u) = -\frac{1}{\,(4\pi)^2\,\alpha\, }\,\hbar\, \dot{S}_{\mathrm{rad}}(u)^2,
\end{equation}
where a dot denotes $\partial_u$. This expression shows that the entanglement entropy of radiation cannot decrease without an energy cost; however, the rate of mass-loss \textit{is} independent of the sign of the rate of change of entanglement entropy.

Therefore, as a quiescence phase $B$ cannot change the entropy (as \eqref{energy_entropy_rel} vanishes), this means that at the beginning of the purification phase $C$ the entanglement entropy of radiation is an enormous excess of entanglement of field modes between the early $(-\infty,0)$ and the late $(0,\tau_C)$ portions of $\scrip$. We require that at the end of phase $C$ the entropy of radiation vanishes,
\begin{equation}\label{LR_EE}
    S_{\mathrm{rad}}(0) = 8\pi \frac{M_0^2}{\hbar}\,,\qquad     S_{\mathrm{rad}}(\tau_C) =0\,. 
\end{equation} 
These boundary conditions, together with \eqref{EE_formula}, impose the constraint
\begin{equation} \label{EE_constraint}
    \int_0^{\tau_C} k(u)\, du = -\frac{2}{\alpha} \frac{M_0^2}{\hbar}\,.
\end{equation}

More generally, if we do not assume that the black hole thermally evaporates until it reaches the Planck mass, so that instead $M(0) \equiv M_C$ is the mass left at the start of the purification phase C, the energy-budget constraint would read $\int_0^{\tau_C} k(u)^2 du = M_{C}/\alpha \hbar$.
Similarly, if the entanglement at the start of the purification phase C is instead left more generally as $S_{\rm rad}(0) \equiv S_{C}$, then the purity-constraint would be $\int_0^{\tau_C} k(u) du = -S_{C}/4\pi \alpha$.
 

\section{Lifetime Bounds}

The constraints \eqref{energy_constraint} and \eqref{EE_constraint} in the purification phase $C$ allow us to put a lower bound on its duration $\tau_C$. In fact, these constraints depend only on the function $k(u)$ in the range $u\in (0,\tau_C)$. They can be written in compact notation as
\begin{align}
\big( k\,,\,k\big)\,=&\;\tfrac{1}{\alpha \sqrt{\hbar}}\,, \label{Fluxconstraint2}\\[.4em]
\big( k\,,\,1\big)\,=&\;-\tfrac{2}{\alpha} \tfrac{M_0^2}{\hbar}\,,\label{EEconstraint2}\\[.6em]
\big( 1\,,\,1\big)\,=&\;\;\tau_C\,,
\end{align}
where we have introduced the inner product
\begin{equation}\label{inner_product}
(f,g) = \int_0^{\tau_{C}} f^*(u) g(u) \,du
\end{equation}
for functions in $L^2([0,\tau_C])$. It follows immediately from the Cauchy-Schwarz inequality 
\begin{equation}
 \big(k\,,1\big)^2 \,\leq\, \big(k\,,k\big) \cdot \big(1\,,1\big)
\end{equation}
that the purification time $\tau_C$ is bounded from below by
\begin{equation} \label{lifetime_bound1}
    \tau_{C} \geq \tau_{min} = \frac{4}{\alpha}\frac{M_0^4}{\hbar^{3/2}}\,.
\end{equation}
We emphasize that, within the conformal-vacuum family relevant for spherically-symmetric evaporation, this bound follows directly from the radiative flux \eqref{rad_flux} and entropy \eqref{EE_formula}, without assumptions about the detailed bulk dynamics during the purification phase C. What this means is that energy conservation and purity demand the lifetime to scale with the initial mass $M_0$ of the black hole \textit{at least} as $M_0^4/\hbar^{3/2}$.

For the more model-agnostic constraints on energy and purification expressed in terms of \(M_C\) and \(S_C\), the minimum lifetime bound becomes
\begin{equation}\label{lifetime_bound1.1}
\tau_C \geq \frac{\hbar S_C^2}{16\pi^2\alpha M_C}.
\end{equation}
In this form, it is clear that the energy-entropy purification-time inequality is model-independent within the conformal-vacuum family, relying neither on bulk physics nor on the specific evaporation paradigm preceding the purification phase. On the other hand, the specific \(M_0^4/\hbar^{3/2}\) bound \textit{does} rely on a specific evaporation paradigm.

It is interesting to note the dependence on $\alpha$ in the lower bound \eqref{lifetime_bound1}-\eqref{lifetime_bound1.1}. In particular, \eqref{alpha} implies that it scales \textit{inversely} with the number of scalar fields, meaning this lower bound likely depends on the field species. This lower bound should not be understood as an \textit{estimate} of the lifetime of a black hole. The full theory of quantum gravity in the bulk, together with a specific initial state, is necessary to predict the quiescence time $\tau_B$  and the purification time $\tau_C$. However, using only energy and entropy considerations, while we cannot put upper bounds, we can find a specific \textit{lower bound} on the lifetime of a black hole. Furthermore, since we have not investigated the fluctuations of the flux, and only its average value, these results only apply as expectation values.


\subsection{Minimal Remnant}

\begin{figure}[t]
\makebox[0pt][c]{%
\includegraphics[width=10.0cm]{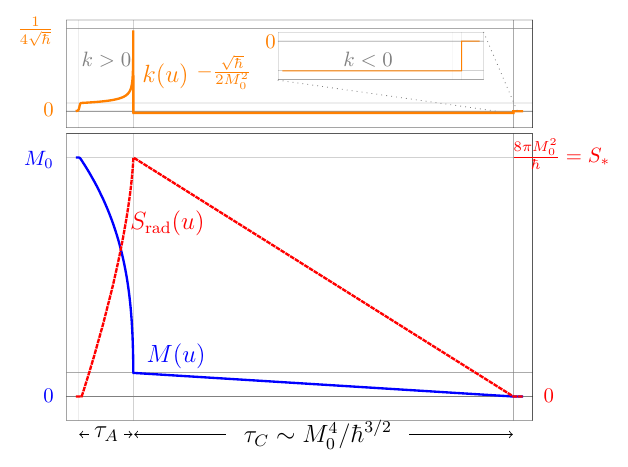}
}
	\caption{\textit{Minimal remnant scenario}. Top: a plot of the redshift exponent $k$ as a function of retarded time $u$. Bottom: plots of the entanglement entropy of radiation $S_{\rm rad}$ and Bondi mass $M$ as a function of retarded time $u$. The purification time is $\tau_C \sim M_0^4/\hbar^{3/2}$.
    }
	\label{fig:Minimal}
\end{figure}

It is interesting to investigate if there exists a profile of the redshift exponent $k(u)$ which saturates the bound \eqref{lifetime_bound1} on the purification phase $C$. Starting with the ansatz $k_C(u)=k_0$ constant over the time $\tau_C$, we have two constraints---energy conservation \eqref{energy_constraint} and purity \eqref{EE_constraint}---which fix the two unknown constants in the ansatz to be
\begin{equation} \label{minimal_kC}
k_C(u)=-\frac{\sqrt{\hbar}}{2\,M_0^2}\,,\qquad \tau_C=\frac{4}{\alpha}\frac{M_0^4}{\hbar^{3/2}}\,.
\end{equation}
This solution of the constraints can be understood as a \emph{minimal remnant scenario}: the black hole of initial mass $M_0$ first evaporates, emitting thermal Hawking radiation until it reaches the Planck mass, and then transitions to a purification phase which lasts the shortest time allowed by the energy budget and purity of the state. 

We note that for an initially large black hole, $M_0\gg \sqrt{\hbar}$, the purification phase lasts much longer than the Hawking thermal phase, $\tau_C\gg \tau_A$. The profile of the redshift exponent, the Bondi mass $M(u)$, and the radiation entropy $S_{\mathrm{rad}}(u)$ are reported in Fig.~\ref{fig:Minimal}. 

It is interesting to observe the change in sign in the redshift exponent $k(u)$. In the Hawking phase $A$, the redshift exponent is positive and given by \eqref{kA}. It depends on a characteristic feature of a semiclassical black hole background: its surface gravity, which is positive. On the other hand, in the purification phase, the redshift exponent turns negative, $k_C(u)<0$. While in this phase we do not have a semiclassical picture of the bulk spacetime, we observe that a negative redshift exponent is a characteristic feature of a white hole. It has been argued that the purification phase represents a black-to-white hole transition \cite{Bianchi:2018mml} (see also \cite{Haggard:2014rza,Christodoulou:2016vny,Christodoulou:2018ryl,DAmbrosio:2020mut,Soltani:2021zmv,Christodoulou:2023psv,Frisoni:2023agk,Dona:2024rdq,Rovelli:2024sjl,Han:2024rqb,Dona:2025snr} and \cite{Ashtekar_2020, Ashtekar_2025}).


\subsection{Metastable Remnant}

\begin{figure*}[t]
	\centering
\includegraphics[width=14.5cm]{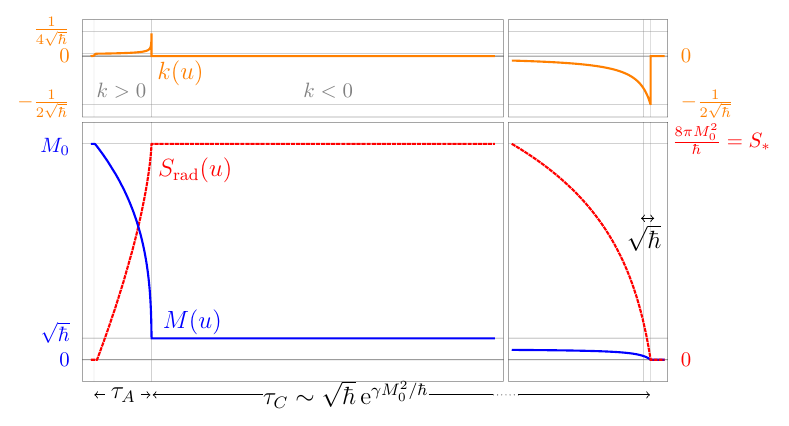}
	\caption{\textit{Metastable remnant scenario}. Top: a plot of the redshift exponent $k$ as a function of retarded time $u$. Bottom: plots of the entanglement entropy of radiation $S_{\rm rad}$ and Bondi mass $M$ as a function of retarded time $u$. The purification time is $\tau_C \sim \sqrt{\hbar}\, \mathrm{e}^{\gamma M_0^2/\hbar}$.
    }
	\label{fig:Metastable}
\end{figure*}

Up to this point, besides the two constraints \eqref{energy_constraint} and \eqref{EE_constraint}, we did not take into account any input from a quantum theory of gravity which describes the quantum geometry in the spacetime bulk. We include now an additional input as motivation: stability of the black hole remnant. Such long-lived remnants have been explored before in the literature from a variety of perspectives \cite{AHARONOV198751, Banks:1992is, Giddings:1993vj, Giddings:1993km, Banks:1994ph, Adler:2001vs, Chen:2014jwq, Christodoulou_2016, Bianchi:2018mml} in the context of information loss. In particular, it is expected that there is a Planck luminosity bound on the energy flux at future null infinity \cite{Wieland:2024kzw}, corresponding to a Planckian Bondi mass. Since adding more flux to decrease the mass at this point would violate this bound, it is possible new physics stabilizes the system to a Planck mass. Note that in loop quantum gravity (see, e.g., \cite{AshtekarLewandowski_LQG_Report, AshtekarBianchi_LQG_Review} for overviews) this is expected  since this is also the point where, due to quantization of the area spectrum \cite{RovelliSmolin1994, RovelliSmolin1995, AshtekarLewandowski_LQGAreaOperator}, the black hole remnant stabilizes as it approaches a Planck area $a_P = \hbar$ gap.

With this in mind, let us define the barycenter of the purification flux as the time
\begin{equation} \label{tau_avg}
    \tau_{\mathrm{bar}} = \frac{\int_0^{\tau_{C}} u\, \langle F_{\mathrm{rad}}\rangle(u) \,du }{\int_0^{\tau_{C}} \langle F_{\mathrm{rad}}\rangle(u\,)  du} \, \equiv \,\alpha \sqrt{\hbar}\int_0^{\tau_{C}}\!\! u \,k^2(u)\, du\,.
\end{equation}
This quantity is the energy-weighted mean emission time during the purification phase and therefore characterizes how long the remnant retains its available energy before releasing it. The condition of metastability can be imposed by maximizing the time $\tau_{\mathrm{bar}}  $ at fixed constraints for energy and purity. We can then determine the profile of the redshift exponent $k(u)$ by solving a simple variational problem. We introduce the functional
\begin{align}\label{functional}
&T[k(u)]\,\equiv\; \alpha \sqrt{\hbar}\,\big(u\,k\,,\,k\big)\,+\nonumber\\[.5em]
&\;\;+\lambda_1\, \bigg( \big (k\,,\,1 \big) + \frac{2}{\alpha} \frac{M_0^2}{\hbar}  \bigg) + \lambda_2\, \bigg( \big (k\,,\, k \big) - \frac{1}{\alpha \sqrt{\hbar}} \bigg).
\end{align}
The first term is the average time $\tau_{\mathrm{bar}} $ \eqref{tau_avg} written in the inner-product notation \eqref{inner_product}. The second and the third terms are the energy constraint \eqref{Fluxconstraint2} and the purity constraint \eqref{EEconstraint2}, with Lagrange multipliers $\lambda_1$ and $\lambda_2$.

The variational principle shows that the extremum for the redshift factor $k(u)$ takes the form
\begin{equation}\label{k-extremum}
\frac{\delta T[k]}{\delta k(u)}=0\;\;\Rightarrow\;\;k(u)=-\frac{1}{a+b\,u}\,.
\end{equation}
Imposing the entropy-constraint \eqref{EE_constraint}, we find a solution for the free parameter $a$ in terms of the second parameter $\tau_C$ and the undetermined constant $b$. Defining $X \equiv \tfrac{2 b M_0^2}{\alpha \hbar}$ for notational convenience, it is given by $a = \tfrac{b \tau_C}{e^X - 1}$. Therefore, we note that $a$ is positive regardless of the sign of $b$.

By taking a second variation of \eqref{functional}, one finds that $\delta^2 T/\delta k^2 = \tfrac{2\alpha \sqrt{\hbar}}{b}(a + b u)$. To understand the possibilities, we note that $a + bu >0$, independently of the sign of $b$: if $b>0$ then this is obvious since $u \in[0,\tau_C]$, but if $b<0$ then we have a linearly-decreasing function of $u$ with endpoints $a>0$ and $a + b\tau_C = a e^X >0$. Thus, the sign of $b$ will completely determine whether or not we have a maximum or minimum: $\textrm{sgn}(b) = \textrm{sgn}(\delta^2 T/\delta k^2)$. In particular, imposing that $b<0$, the flux-barycenter is maximized, i.e., the remaining radiation is released only at the very end of its exponential lifetime \eqref{lifetime_bound2}.

Next, using this value of $a$ and the profile \eqref{k-extremum} in the energy-constraint \eqref{energy_constraint}, we find the free parameter $\tau_C$ in terms of the constant $b$. Keeping track of the first sub-leading order in the large parameter $M_0/\sqrt{\hbar}\gg1$, we get
\begin{equation} \label{lifetime_bound2}
    \tau_{C} = \tfrac{\alpha}{ b^2}  \sqrt{\hbar}\,\exp\Big({\frac{2 |b|}{\alpha}\frac{M_0^2}{\hbar}}\Big)\;-\,\tfrac{2\alpha}{ b^2}  \sqrt{\hbar},
\end{equation}
and therefore the redshift profile that maximizes the flux-barycenter is
\begin{equation}\label{k_C}
    k_C(u) = -\frac{1}{|b|} \frac{1}{(\tau_C + \frac{\alpha}{b^2}\sqrt{\hbar}) - u}\,.
\end{equation}

This solution of the constraints can be understood as a \emph{metastable remnant scenario} for black hole evaporation: the black hole of initial mass $M_0$ first evaporates, emitting thermal Hawking radiation until it reaches the Planck mass, and then transitions to a purification phase which is metastable, and lasts the longest time allowed by a purely-purifying phase $C$, constrained only by the energy budget and purity of the state. Note however, this exponential lifetime is not an absolute upper bound since the variational problem cannot fix the value of $|b|$. As with the lower bound \eqref{lifetime_bound1}, this formula is also sensitive to the number of scalar fields due to the presence of $\alpha$ in \eqref{lifetime_bound2}.

We note that for an initially large black hole, $M_0\gg \sqrt{\hbar}$, the purification phase lasts an exponentially long time, $\tau_C\sim \sqrt{\hbar}\, \mathrm{e}^{\gamma M_0^2/\hbar}$. The quantity in the exponential is proportional to the Bekenstein-Hawking entropy $A_0/4\hbar$ \cite{Hawking:1974rv, Bekenstein73, Hawking1976_BHT, AshtekarParaizoShu-PRL, AshtekarParaizoShu-GRG} of the initial black hole. The profile of the redshift exponent, the Bondi mass $M(u)$ and the radiation entropy $S_{\mathrm{rad}}(u)$ for this scenario are reported in Fig.~\ref{fig:Metastable} for the case $b<0$. 

We observe again the change in sign in the redshift exponent $k(u)$ in the purification phase, $k_C(u)<0$. While describing the bulk spacetime in this phase requires a full theory of quantum gravity, at future null infinity we can rely on asymptotic quantization and characterize the phase purely in terms of the redshift exponent $k(u)$, which is known to be negative for a white hole. Therefore, the metastable remnant scenario is suggestive of a transition from a Planck mass black hole to a Planck mass white hole \cite{Bianchi:2018mml}. Eventually, after a time $\tau_C$, the white hole completely decays, emitting a burst of purifying radiation, with $k_C(u\rightarrow \tau_C) \sim -1/\sqrt{\hbar}$ going to zero in a Planck time $t_P=\sqrt{\hbar}$. Hawking's thermal black hole explosion at the time $\tau_A$ \cite{Hawking:1974rv} is followed, after an exponentially long time in the initial horizon area $A_0\sim M_0^2$, by a purifying white hole implosion at the time $\tau_C$. In order for the flux at the end of the purifying phase to satisfy the Planck luminosity bound \cite{Wieland:2024kzw}, we would need $b^2 \sim \alpha$. This physical condition could be used to fix the otherwise undetermined coefficient of the exponential lifetime \eqref{lifetime_bound2}.


\section{The Moving Mirror Analog}

In the previous sections, we adopted the perspective of asymptotic quantization, relying only on the structure of $\scrip$ -- where physics is expected to remain semiclassical -- without any reference to the bulk spacetime, which is expected to be far from semiclassical in the purifying phase. Nevertheless, it is useful to provide a picture of an effective bulk spacetime inferred from the properties of the correlation functions at $\scrip$ and specifically the redshift factor $k(u)$. It is well-known that the $(3+1)$d spherically-symmetric model considered here can be described in terms of a moving mirror analog that effectively `fills in' the bulk of the evaporating black hole \cite{Davies-Fulling_MM, Davies-Fulling_MM+BH, Davies-Fulling76,Varadarajan-2025} (see also \cite{Wald_2019, Walker1985, Carlitz-Wiley1987_Reflections, FordRoman_MM_Neg_Energy, wilczek1993, Ford_2004, Good_2013, Good_2015, Good2016, Chen_2017, Good2018, Good2019, Tomitsuka_2020, Agullo_2024, Good:2026dmf}).

In this section we translate the results found above into the language of the moving mirror model. In this model, the ray-tracing function $v=p(u)$ describes how a radial light ray incoming from the advanced time $v$ at $\scrim$ emerges at $\scrip$ at the retarded time $u$. In double-null coordinates $(u,v,\theta,\phi)$, the ray-tracing function describes the axis of spherical symmetry $r(u,v)=0$. Equivalently, the function $v=p(u)$ can be understood as the trajectory of a moving mirror in an auxiliary $2$d Minkowski spacetime. The redshift factor for radial null rays then becomes simply the redshift factor $\dot{p}(u)\equiv\partial_u p(u)$ for light rays reflecting off the moving mirror. Its logarithmic derivative, 
\begin{equation}
\label{eq:ku}
-\ddot{p}(u)/\dot{p}(u) \equiv k(u)\,,
\end{equation}
is precisely the redshift exponent $k(u)$, also known as the peeling function \cite{Barcel2011-HawkRad, Barcel2011-Minimal_Conditions, Agullo_2024, Agullo:2026}. Moreover, the correlations in the in-vacuum $\ket{0_{\mathrm{in}}}$ are $\langle 0_\mathrm{in}|\partial_u\chi_0(u)\,\partial_{u'}\chi_0(u')|0_\mathrm{in}\rangle=-\frac{1}{4\pi}\frac{\hbar}{4\pi}\frac{\dot{p}(u)\dot{p}(u')}{(p(u)-p(u'))^2}$ and the renormalized expectation value, which defines the stress-tensor flux, is simply given by $\langle \partial_u\chi_0\,\partial_u\chi_0\rangle(u)\,=\tfrac{1}{4\pi}\;\tfrac{\hbar}{48\pi}\big(k^2(u)+2\dot{k}(u)\big)$ in terms of the redshift exponent $k(u)$ \cite{BianchiParaizo2026,Holzhey_1994, wilczek1993}. We note that this is \textit{not} the same as the radiative flux \eqref{rad_flux} identified in \cite{BianchiParaizo2026}. The proper identification of the radiative flux naturally absorbs the `Coulombic' $\dot{k}-$term into the mass \eqref{mass-balance}. We refer the reader to \cite{BianchiParaizo2026} for further details.

\begin{figure}
  \centering
  \includegraphics[scale=0.65]{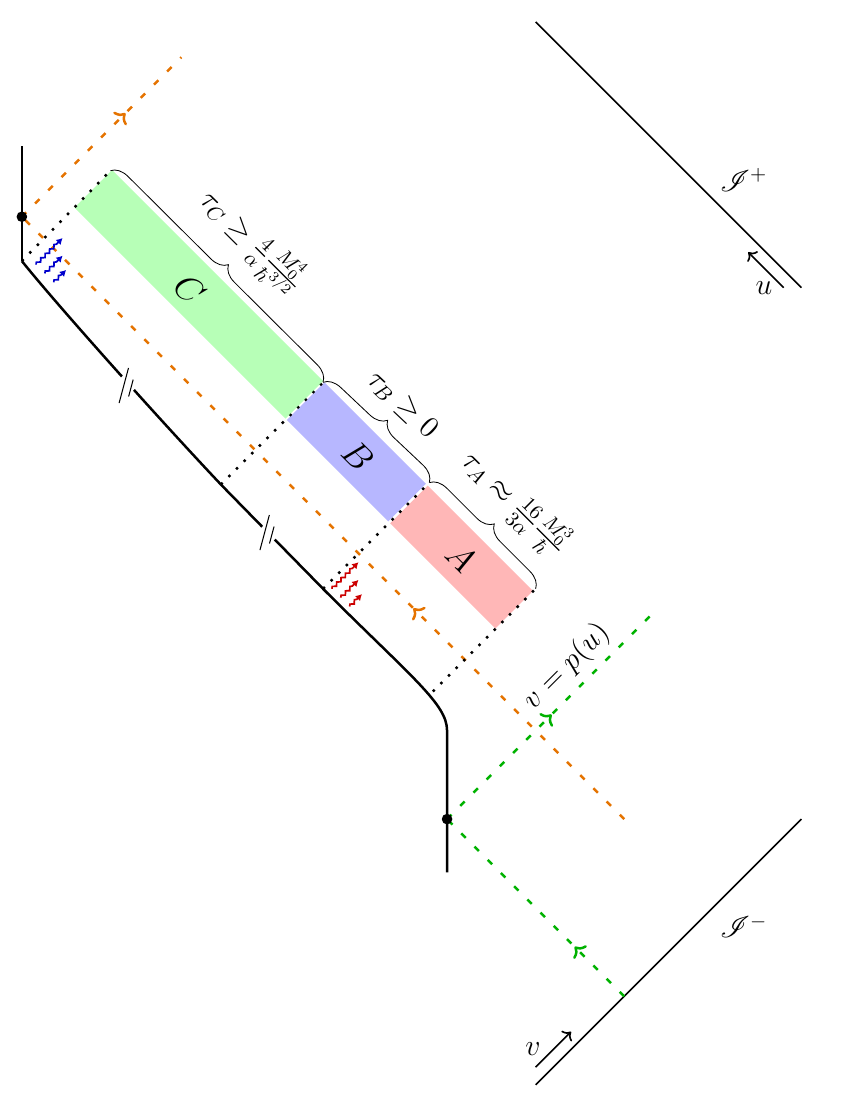}
  \caption{The moving mirror trajectory for the bulk spacetime motivated from our asymptotic considerations for spherically-symmetric black hole evaporation. The ray tracing function $p(u)$ determines the trajectory. It is thermal during phase A and highly constrained during phases B, C, and afterwards.}
  \label{mirror}
\end{figure}

In this interpretation, the phases $A$, $B$, and $C$ take on a clear physical picture, as  described for instance in \cite{wilczek1993}. Phase $A$ corresponds to a mirror that is accelerating, asymptotically approaching a null trajectory. During this phase, $p(u)$ takes on the well-known thermal form (see, for instance, \cite{Agullo_2024, Agullo:2026}). Phase $B$ is when the mirror decelerates and becomes inertial, no longer emitting radiation. Finally, phase $C$ is when the mirror transitions to a final inertial trajectory, emitting the purifying partners of the earlier radiation. During this phase, the mirror trajectory can be obtained by twice integrating the redshift exponent $k_C(u)$ in \eqref{minimal_kC} or \eqref{k_C}. Fig.~\ref{mirror} provides a depiction of these three phases.

We recall that the energy flux vanishes in phase $B$ and after phase $C$. As a result, from \eqref{rad_flux} it immediately follows that when this occurs, the ray-tracing function becomes linear: $p(u) = a u + b$. This is a two-parameter family of trajectories, i.e. the mirror returns to an inertial trajectory, but is not necessarily at rest with respect to the initial frame. Imposing only the condition of vanishing flux, we cannot constrain the constant $a$. Thus, during phase $B$, the mirror can be highly boosted with respect to the initial rest frame. Apriori, the same could be true when the flux vanishes after phase $C$; but as we will show below, this is not the case.

To see this, we note that the entanglement entropy \eqref{EE_formula} can be equivalently written using \eqref{eq:ku} as $S_{\mathrm{rad}}(u) = -4\pi \alpha \log{\dot{p}(u)}$. The requirement that the entanglement entropy drops to zero at the end of the purifying phase $C$, i.e. $\lim_{u \ra \tau_C} S_{\mathrm{rad}}(u)=0$, imposes the condition
\begin{equation} \label{zero_entropy}
   \lim_{u \ra \tau_C} \log{\dot{p}(u)} = 0.
\end{equation}
This occurs for any $p(u)$ of the form
\begin{equation} \label{ray-tracing}
    p(u) = u + b.
\end{equation}
Therefore, the mirror returns to rest, but is displaced by the amount $b$. Since the ray-tracing function $p(u)$ completely determines the causal structure of the effective spacetime, we see that these simple considerations fix the global causal structure following complete evaporation at the end of phase $C$. This is in-line with the arguments in \cite{Varadarajan-2025, Bianchi:2018mml}. Furthermore, this result provides a more stringent constraint: \textit{there is now only one free parameter $b$, a `memory', for the ray-tracing function in the far future after phase $C$.} 

In summary, the formulas for the radiative energy flux \eqref{rad_flux} and entanglement entropy of radiation \eqref{EE_formula} of a spherically-symmetric evaporating black hole in four dimensions can be used to completely fix the qualitative profile of the mirror trajectory analog of black hole evaporation. The Hawking phase $A$ is thermal. The quiescent phase $B$ is inertial but boosted with respect to the initial frame. The phase $C$ is obtained by integrating \eqref{minimal_kC} or \eqref{k_C}, and complete evaporation results in the mirror returning to an inertial trajectory that is at rest with respect to the initial frame, but with a  `memory' given by the constant $u$-shift $b$.


\section{Discussion}

In this paper, we determined constraints on the lifetime of a black hole which follow from the requirement of energy conservation and unitary evolution. We worked in a $3+1$ spherically-symmetric model of black hole evaporation and connected it to results in the moving mirror model. The analysis uses prior results on the entanglement entropy of radiation at $\scrip$ found in \cite{Bianchi:2014bma, Bianchi:2014qua, Bianchi:2014vea} and on the radiative flux of Hawking radiation from \cite{BianchiParaizo2026}. These formulas, and therefore the results in this work, rely on a reduction of the relevant quantum states -- provided by the spherically-symmetric model -- to conformal vacua. We are led to the scenarios depicted in Fig.~\ref{penrose_diagram}--\ref{mirror}. We conclude with a few points of discussion.

\begin{itemize}[leftmargin=1.2em,label=\raisebox{.14em}{$\scriptstyle\bullet$}]

    \item As stated in section II, this work presupposes that (i) quantum gravitational effects in the bulk do not impede the validity of semiclassical theory at null infinity; and (ii) information is recovered in the Page-paradigm sense, i.e., that the entanglement entropy of radiation starts and ends at zero.
    
    The first is a reasonable assumption on the nature of quantum gravity, in that an asymptotic, weak-field regime of low-curvature continues to exist. It is the assumption of `asymptotically semiclassical spacetimes'. Within this conceptual framework, one example of an obvious `bad' quantum gravity effect we are discarding is that of naked and thunderbolt singularities \cite{Hawking_1993}. If either of these generically occur at the tail-end of evaporation, evolution beyond the last ray would simply not be possible. In that sense, spacetime would come to an end, as divergent curvature would be present infinitely far away. One would completely lose the notion of that black hole spacetime being asymptotically flat, and more importantly, one would lose predictive power. But already in the detailed semiclassical analysis of 2d dilatonic black holes \cite{Callan_1992} done by Ashtekar-Taveras-Varadarajan (ATV), this was not found to occur \cite{Ashtekar:2008jd, Ashtekar:2010hx, Ashtekar:2010qz}. Notice that this is \textit{not} the same as requiring small fluctuations in the stress-tensor flux \cite{FordKuo_Semicl, Wu_1999, Ford:2000vm}, which we never imposed or investigated. Indeed, one can have large fluctuations for reasonable states and spacetimes -- even squeezed states in Minkowski -- where one does not expect semiclassical theory to fail; in other words, demanding curvature to remain small asymptotically is a significantly weaker condition than that of small flux-fluctuations. Since we do not control these fluctuations, this is why we emphasized in section V that our results are not sharp estimates, but only conditions on the expected values.

    The second assumption presupposes not just that Hawking radiation at $\scrip$ will be purified by correlations, but that these correlations will be accessible to a fiducial asymptotic observer. Note that this assumption and the considerations in this paper are completely independent of what happens near the dynamical horizon where the curvature becomes Planckian in the late stages of the evaporation. But it does mean we are excluding the existence of an \emph{event} horizon and of any kind of `baby universe' scenario, in which the final state is pure---but the purifying correlations remain inaccessible (see, e.g., \cite{Unruh_2017, Marolf_2017, Hossenfelder_2010}).

    \item The energy-entropy relation \eqref{energy_entropy_rel} indicates that purification cannot occur at zero energy-cost \cite{Hotta_PartnerParticles, Wald_2019, Agullo:2026}. Indeed, it implies that once the flux goes to zero, the entanglement entropy does not change. However, if the mass at the start of purification is the Planck mass \eqref{eq:planck-mass}, then the large entanglement at the last ray \eqref{LR_EE} indicates that purification would not yet have occurred. Additionally, this large entanglement entropy (of the Hawking radiation at $\scrip$), combined with the small mass of the black hole at the last ray, indicates a state similar to some kind of `remnant' \cite{AHARONOV198751, Banks:1992is, Giddings:1993vj, Giddings:1993km, Banks:1994ph, Adler:2001vs, Chen:2014jwq, Christodoulou_2016, Bianchi:2018mml}.
    
    \item We directly gave a one-line proof for the $M_0^4/\hbar^{3/2}$ lifetime-bound \eqref{lifetime_bound1} by simply combining the formulas for the flux and entropy. Figure \ref{fig:Minimal} depicts the behavior of the redshift exponent $k(u)$, Bondi mass $M(u)$, and the entanglement entropy $S(u)$ (i.e, Page curve). Such a bound is consistent with what has been explored before in \cite{Carlitz-Wiley1987_Lifetime, preskill1992blackholesdestroyinformation, phdthesis} (see also \cite{Martin-Dussaud:2025qtr}), but these works had to rely on extra assumptions about the dynamics in the bulk. This bound is also in-line with the analysis of \cite{AlmheiriandSully} that investigated the entanglement entropy of 2d dilatonic black holes \cite{Callan_1992} using the ATV flux-proposal \cite{Ashtekar:2008jd,Ashtekar:2010hx,Ashtekar:2010qz,Barenboim:2025fds}. However, since these black holes do not have the correct heat capacity (as their temperature does not change with the mass), there did not appear the correct mass-scaling in the entropy (in particular, it was found in \cite{AlmheiriandSully} that $S(u_{\rm lr})$ was linear in $M$, rather than quadratic).

    \item With a \textit{simple additional assumption}, we showed that the exponential lifetime \eqref{lifetime_bound2} on the lifetime arises. The additional assumption, which relies on quantum gravity considerations, is that a Planck-area black hole is metastable: it retains its remaining mass for as long as possible before releasing it near the end of the purification phase. We determined the configuration that maximizes the barycenter of the purifying flux, which is depicted in the plots of $k(u)$, $M(u)$, and $S(u)$ in Fig.~\ref{fig:Metastable}. See also the  analysis in \cite{Agullo:2026} and \cite{Ho:2026xhu} that argued for such a bound on different grounds.

    \item The current analysis cannot predict how long the transition phase $B$ lasts. In particular, we cannot provide \textit{upper} bounds on lifetime. Furthermore, at the end of both phase $A$ and phase $C$, the redshift exponent $k(u)$ takes on a large, Planckian value (i.e., $k\propto 1/\sqrt{\hbar}$) -- which satisfies the Planck luminosity bound \cite{Wieland:2024kzw} -- before then dropping to zero. This is clearly visible in Fig.~\ref{fig:Minimal}--\ref{fig:Metastable}, and is an extreme transition in $k(u)$. Additionally, the metastability condition results in a lifetime that is exponential in the square of the initial black hole mass, i.e. in its initial area. During this purification phase, we find that the redshift exponent is negative, which indicates the existence of a white-hole remnant \cite{Bianchi:2018mml,Haggard:2014rza,Christodoulou:2016vny,Christodoulou:2018ryl,DAmbrosio:2020mut,Soltani:2021zmv,Christodoulou:2023psv,Frisoni:2023agk,Dona:2024rdq,Rovelli:2024sjl,Han:2024rqb,Dona:2025snr}. These transitions in $k(u)$ possibly indicate highly non-trivial effects arising from quantum gravitational processes in the bulk, which require a full theory of quantum gravity to be addressed.

    \item Primordial black holes produced in a hot cosmic environment with temperature $T\approx 10^{9}\,\mathrm{GeV}$ have an initial mass $M_0\sim \hbar^{3/2}/T^2\,\approx 10^{12}\,\mathrm{kg}$. Their Hawking thermal phase lasts $\tau_A\sim M_0^3/\hbar\,\approx 1$~billion years, and the remnant purifying phase is much longer than the age of the universe, $\tau_C>  \tau_{\mathrm{min}}\sim M_0^4/\hbar^{3/2}\sim 10^{20}\, \tau_A$. 
    
    On the other hand, for primordial black holes in the mass range $M_0\approx 10^3\,\mathrm{kg}$ (see \cite{Vidotto:2026pbh}), the Hawking thermal explosion happens in a short time $\tau_A\sim M_0^3/\hbar\,\approx 10^{-11}\,\mathrm{s}$. Once they reach the Planck mass, they enter the purifying phase. The minimal bound for the purifying phase is a short $\tau_{\mathrm{min}}\sim M_0^4/\hbar^{3/2}\,\approx 1\,\mathrm{s}$. However, when the metastability condition is taken into account, the metastable scenario (Fig.~\ref{fig:Metastable}) predicts remnants that last an exponentially long time $\tau_C\sim \sqrt{\hbar}\,\mathrm{e}^{\gamma M_0^2/\hbar}$, which is much longer than the age of the universe. They are therefore practically stable Planckian remnants. We refer to \cite{Vidotto:2026pbh} for a recent study of the phenomenological implications of stable primordial black holes remnants in this mass range.

\end{itemize}


\bigskip

\emph{Acknowledgments}. We thank Francesca Vidotto, Carlo Rovelli, Wolfgang Wieland, and Ivan Agullo for insightful discussions. E.B. is especially grateful to Matteo Smerlak, Tommaso De Lorenzo, and Pierre Martin-Dussaud, for many discussions on information and the lifetime of a black hole. We also thank an anonymous referee for carefully reading the manuscript and providing helpful suggestions. D.E.P. acknowledges support via the Bunton-Waller award from Penn State and the Walker fellowship from the Applied Research Lab. E.B. is supported by the National Science Foundation, Grants No. PHY-2207851 and PHY-2513194. This work was made possible thanks to the support of the WOST project (\href{https://withoutspacetime.org}{\mbox{withoutspacetime.org}}), funded by the John Templeton Foundation (JTF) under Grant ID 63683. 


\bibliography{references}

\end{document}